\documentclass[sigconf,screen,nonacm,table]{acmart}

\renewcommand\footnotetextcopyrightpermission[1]{}
\setcopyright{none}
\usepackage{multirow}  
\usepackage{booktabs}  
\usepackage{enumitem}
\usepackage{booktabs}
\usepackage{tabularx} 
\usepackage{graphicx}
\setkeys{Gin}{draft=false}
\usepackage{algorithm}
\usepackage{algpseudocode}

\begin{document}

\title{Gated Coordination for Efficient Multi-Agent Collaboration in Minecraft Game}

\author{%
HuaDong Jian, Chenghao Li, Haoyu Wang, Jiajia Shuai, Jinyu Guo, Yang Yang, Chaoning Zhang$^{*}$%
}

\affiliation{%
  \institution{University of Electronic Science and Technology of China}
  \city{Chengdu}
  \country{China}
}

\renewcommand{\shortauthors}{Jian et al.}

\begin{abstract}
In long-horizon open-world multi-agent systems, existing methods often treat local anomalies as automatic triggers for communication. This default design introduces coordination noise, interrupts local execution, and overuses public interaction in cases that could be resolved locally. To address this issue, we propose a partitioned information architecture for MLLM agents that explicitly separates private execution states from public coordination states.
Building on this design, we introduce two key mechanisms. First, we develop an event-triggered working memory based on system-verified outcomes to maintain compact and low-noise local state
representations. Second, we propose a cost-sensitive gated escalation mechanism that determines whether cross-region communication should be initiated by jointly considering node criticality, local recovery cost, and downstream task impact. In this way, communication is transformed from a default reaction into a selective decision. Experiments conducted on long-term construction tasks in open environments demonstrate that, compared to baseline models based on strong communication and planned structures, the
introduction of gated communication and a partitioned information architecture results in superior performance in terms of blueprint completion quality and execution chain length. It also improves local self-recovery, reduces ineffective escalations, and increases the utility of public communication.
\end{abstract}

\keywords{MLLM Agents, Multi-Agent Systems, Agent Coordination, Long-Horizon Planning, Open-World Environments}

\begin{teaserfigure}
    \centering
    \includegraphics[width=0.88\textwidth,trim=0 2.5cm 0 2cm,clip]{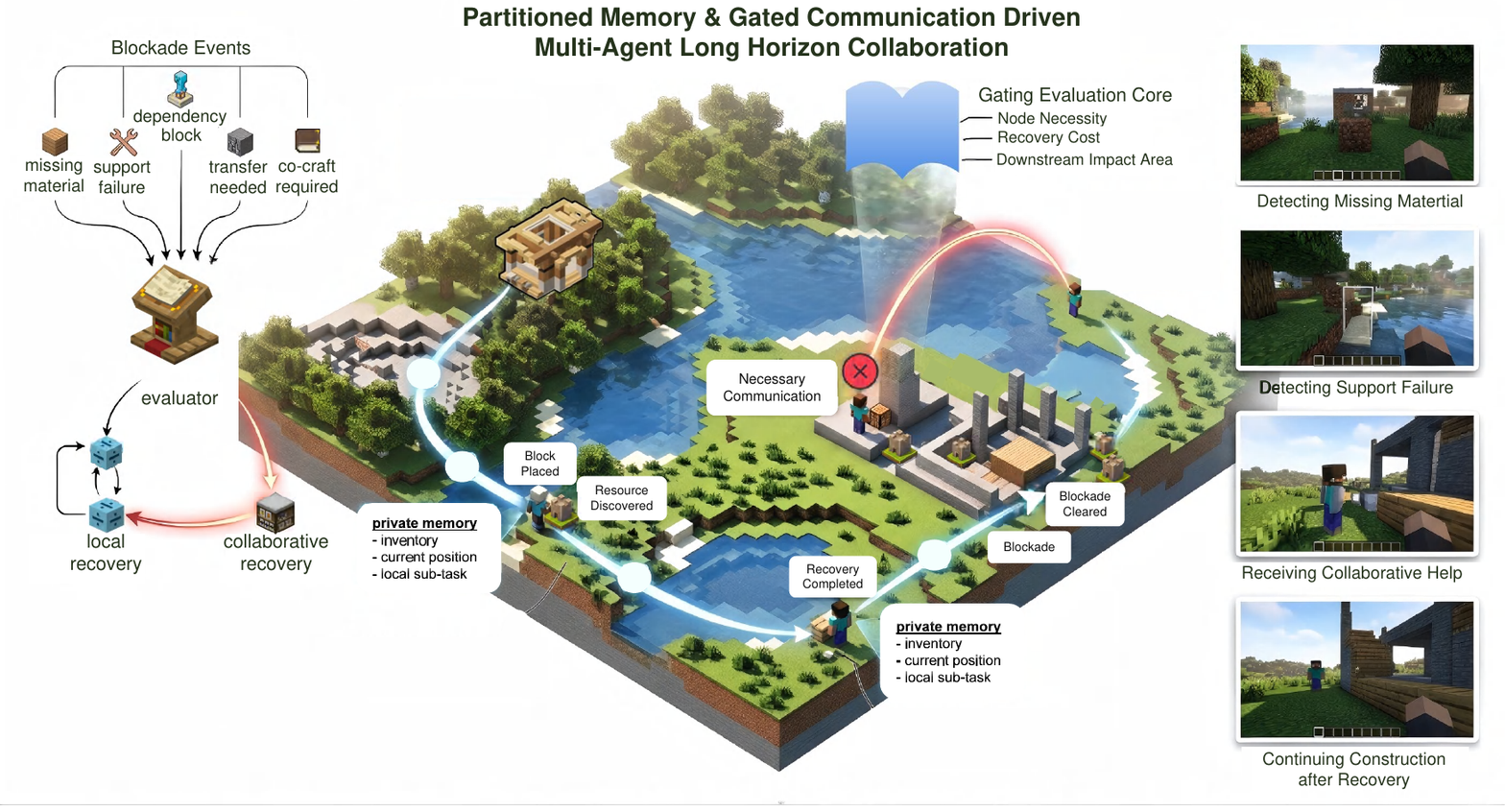}
    \caption{Illustration of our gated collaborative escalation framework in a long-horizon Minecraft construction task. Agents primarily operate within private memory-driven local execution loops, while structured blockade events trigger a gated evaluation process that decides whether to stay local or escalate into public coordination. Verified action outcomes are consolidated back into private memory, and communication is opened only when collaboration is necessary for efficient recovery and continued progress.}
    \label{fig:teaser}
\end{teaserfigure}
\maketitle

\section{Introduction}
\label{sec:introduction}

As multimodal large language models (MLLMs) continue to improve their capabilities in planning, tool use, and environmental interaction \citep{zhai2025survey,xu2025deploying,zhao2024see,huang2022language,song2023llm,driess2023palm,zitkovich2023rt}, MLLM-based multi-agent systems are rapidly emerging as a key paradigm for solving tasks in open-ended environments \citep{chen2023agentverse,hong2023metagpt,qian2024chatdev,shaikh2025llm,huang2025scaling}. In complex domains like Minecraft, characterized by open-world attributes, intricate resource dependencies, long execution cycles, and multi-layered task chains---a single agent often struggles to accomplish overarching objectives independently \citep{zhao2024see,fan2022minedojo,wang2023voyager,lifshitz2023steve,qin2024mp5,wang2024omnijarvis}. Consequently, enhancing task completion efficiency through role division, resource sharing, and dynamic collaboration among multiple agents has become a prominent research direction \citep{wu2024autogen,white2025collaborating,sun2025collab,chai2025causalmace}.

Existing MLLM-based multi-agent methodologies heavily emphasize the benefits of such collaboration \citep{li2023camel,hong2023metagpt,qian2024chatdev,shaikh2025llm}. Generally, these approaches fall into two categories: treating communication as a continuous channel for context sharing and plan synchronization \citep{wu2024autogen,white2025collaborating}, or binding communication to specific event triggers (e.g., resource shortages or path blockages) to ensure timely responses to anomalies \citep{shaikh2025llm,huang2025scaling,chai2025causalmace}. Despite their algorithmic differences, both lines of work converge on a shared, implicit assumption: "that more communication strictly leads to better coordination". They operate under a "communication-first" paradigm, presuming that once local uncertainty arises, immediate cross-agent interaction is universally preferable to delayed, local handling \citep{hong2023metagpt,qian2024chatdev,huang2025scaling}.

However, while this premise may hold in short-horizon tasks, it becomes increasingly brittle in long-horizon settings \citep{wang2023voyager,shaikh2025llm,xi2025agentgym}, yielding three detrimental consequences. First, it \textbf{interrupts continuous local execution}. Unregulated communication is not a zero-cost exchange; it consumes cognitive budgets and action windows \citep{zhai2025survey,zhang2024cut,li2025parallelized}, forcing agents to suspend ongoing local plans. Second, it \textbf{amplifies global state-update noise}. Frequent interaction often prematurely escalates minor local deviations into system-wide coordination events, triggering cascading interference across agents \citep{huang2025scaling,chen2025optima,xu2026mitigating}. Third, it \textbf{leads to false coordination triggers}. Current mechanisms are predominantly driven by the mere presence of an anomaly, devoid of cost-benefit analysis \citep{shaikh2025llm,huang2025scaling,chen2025optima,zhang2024cut}. By inherently understating an agent's local self-repair capabilities (e.g., alternative resource gathering or localized re-planning), these systems erroneously conflate the ability to communicate with the necessity to do so \citep{zhao2024see,wang2023voyager}.

Consequently, achieving true collaborative efficiency in long-horizon scenarios requires a paradigm shift. The critical research gap is not the need for more sensitive communication triggers, but rather the absence of \textbf{a fine-grained adjudication layer}. Communication should not serve as the default successor action following an anomaly; it must be the calculated result of a tradeoff between autonomous local handling and collaborative escalation. Existing systems have yet to answer a pivotal question: which issues warrant escalation to public coordination, and which should remain confined to private local execution? Without this capability, current methods achieve superficial responsiveness at the severe expense of global execution stability and communication efficiency\citep{zhang2024cut,chen2025optima,xu2026mitigating}.

To bridge this gap, we propose a \textbf{Partitioned Information Architecture} that explicitly separates the private execution state from the public coordination state in MLLM-driven multi-agent systems. Each agent maintains a deterministic, private self-perception zone that stores only execution-critical information, serving as a compact local working memory updated exclusively via system-verified outcomes. Concurrently, a public communication zone strictly contains state-changing coordination signals. To couple these layers seamlessly, we introduce a \textbf{Gated Escalation Mechanism} that authorizes cross-zone communication only when predicted to be globally beneficial. By comprehensively evaluating node criticality, local recovery costs, and downstream task impacts, this design empowers agents to preserve stable local execution while exposing private states for coordination only when strictly necessary.

We evaluate the proposed system in both standard and resource-constrained custom construction tasks. Experimental results demonstrate that our method significantly outperforms existing baselines in robustness and execution efficiency. Notably, in complex custom scenarios where standard models suffer severe performance degradation due to coordination deadlocks and noise, our approach reduces the total number of completion steps while maintaining a high task success rate. By explicitly decoupling private and public states, our framework fundamentally reshapes coordination dynamics, transforming communication from a default reflex into a selective decision. This ensures public interactions have clear objectives, achieving an optimal balance between task success and minimal communication overhead.

Our main contributions are as follows:

\begin{enumerate}[leftmargin=*]

    \item We propose a novel framework for MLLM agents that explicitly decouples private execution states from public coordination states, minimizing contextual noise and global disruption.

    \item We introduce a gating mechanism that controls cross-zone access, restricting inter-agent interaction to necessary, state-changing coordination based on rigorous cost-benefit evaluations, thereby drastically improving execution robustness. 

    \item We design an event-triggered, private working memory that maintains compact and verifiable local states using strictly system-verified action outcomes, circumventing the hallucination risks of free-form MLLM summarization.

\end{enumerate}
\section{Related Work}
\label{sec:related_work}

\subsection{MLLM Agents in Open Worlds}
Standardized by benchmarks like MineDojo \citep{fan2022minedojo,lin2023mcu,mohanty2025idat}, open-world environments require agents to seamlessly integrate long-horizon planning with dynamic execution. Existing research primarily advances agent capabilities along three dimensions \citep{huang2022language,song2023llm,wang2023voyager,yuan2023skill,zhu2023ghost,wang2024jarvis,qin2024mp5,wang2024omnijarvis,wang2023describe}: \textbf{(1)} Hierarchical Planning: Voyager \citep{wang2023voyager} introduces an iterative programmatic skill library for continual learning, while Plan4MC \citep{yuan2023skill} adopts a two-tier paradigm decoupling top-level LLM planning from low-level skill execution.
\textbf{(2)} Memory Integration: To navigate vast state spaces, GITM \citep{zhu2023ghost} and JARVIS-1 \citep{wang2024jarvis} leverage external knowledge bases and multimodal memory to support extended material gathering and crafting. \textbf{(3)} Error Recovery: Frameworks like DEPS \citep{wang2023describe,yao2022react,liu2025survey} utilize self-reflection to trigger global replanning after execution failures. Despite these advances, most approaches implicitly assume deterministic underlying environments, relying heavily on post-hoc recovery only after a complete task failure \citep{wang2023voyager,zhu2023ghost,wang2024jarvis,wang2023describe,qin2024mp5,wang2024omnijarvis,yu2024adam}. Mechanisms for systematically absorbing continuous, micro-level local execution uncertainty on the fly---without derailing global task progress—remain largely underexplored.

\subsection{MLLM-powered Multi-Agent Collaboration}
To overcome the cognitive and operational bottlenecks of single agents, multi-agent frameworks predominantly focus on structuring communication topologies and interaction protocols \citep{li2023camel,chen2023agentverse,hong2023metagpt,qian2024chatdev,wu2024autogen,chen2023autoagents,white2025collaborating,li2025parallelized,shaikh2025llm}. Organizationally, prior works often employ explicit hierarchical or graph-based task synchronization \citep{chen2024s,dong2024villageragent,hong2023metagpt,wu2024autogen}: HAS \citep{zhao2024hierarchical} utilizes a star topology for centralized planning and decentralized execution; S-agents \citep{chen2024s} designs an asynchronous agent tree; and VillagerAgent \citep{dong2024villageragent} implements a strict top-down, DAG-based planner. Regarding proactive interaction, ProAgent \citep{zhang2024proagent} models teammate intent for dynamic decision-making, whereas CoELA \citep{zhang2023building} uses memory retrieval to drive structured natural language communication. However, current paradigms typically treat communication as a reactive, event-driven mechanism rather than a proactive decision \citep{wu2024autogen,zhang2024proagent,white2025collaborating,chai2025causalmace}. As highlighted by Mindcraft \citep{white2025collaborating}, inefficient information exchange remains a core bottleneck. The field lacks fine-grained communication policies that perform a cost-benefit evaluation, weighing the effort of independent local resolution against the utility of querying teammates---to prevent cognitive overload and foster genuinely efficient collaborative dynamics \citep{chen2025optima,zhang2024cut,yue2025masrouter,li2025parallelized}.
\begin{figure}[t]
    \centering
    \includegraphics[width=1.04\linewidth,trim=0  1.8cm 0 6cm,clip]{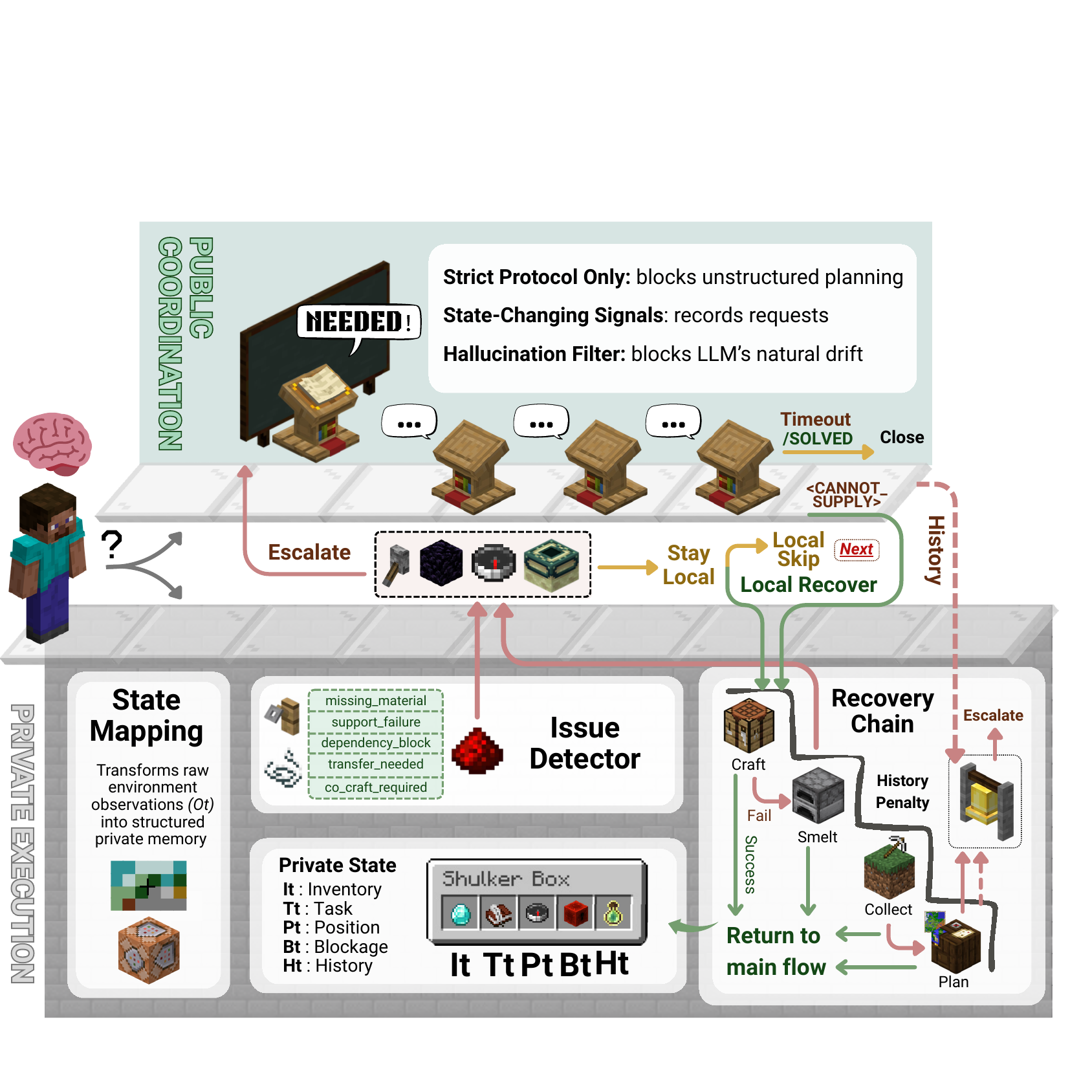}
    \vspace{-2mm}
    \caption{Overview of the partitioned information architecture coupled with a gated escalation mechanism.}
    \label{fig:my_image}
    \vspace{-2em} 
\end{figure}

\section{Methodology}

\label{sec:method}

\subsection{Method Overview}
To operationalize the principle that communication is a costly coordination action rather than a default reflex \citep{chen2023agentverse,white2025collaborating,chen2025optima,zhang2024cut,li2025parallelized,yue2025masrouter}, we model cross-agent interaction as a deliberate, gated escalation process. While traditional systems eagerly broadcast local anomalies-frequently resulting in state pollution, execution interruptions, and coordination deadlocks \citep{chen2023agentverse,dong2024villageragent,chen2024scalable,white2025collaborating,chen2025optima,chai2025causalmace}, our framework strictly confines absorbable issues to the local level.

Our objective is not to suppress collaboration entirely, but to render it \textit{selective, cost-sensitive, and reversible}. Specifically, we design the system to satisfy three core desiderata:
\begin{itemize}[leftmargin=*]
    \item \textbf{Preserve local execution continuity:} Ensure that locally resolvable issues form a closed-loop within the execution layer to minimize disruption.
    \item \textbf{Restrict public coordination:} Expose private states to the public channel only when collaboration yields explicit, quantifiable benefits.
    \item \textbf{Tolerate coordination failures:} Equip the system with graceful fallback mechanisms (e.g., local recovery or skipping) to avert infinite waiting states upon collaboration failure.
\end{itemize}

To achieve this, we propose a \textbf{partitioned information architecture coupled with a gated escalation mechanism}. Formally, we define the control flow of a single agent at time step $t$ as the following decision chain:
\begin{equation}
o_{t} \rightarrow m_{t}^{\text{priv}} \rightarrow d_{t}^{\text{issue}} \rightarrow g_{t}^{\text{esc}} \rightarrow a_{t}
\end{equation}
where $o_{t}$ denotes the current environmental observation and system feedback; $m_{t}^{\text{priv}}$ is the updated private execution state; $d_{t}^{\text{issue}}$ represents the detection result of structured construction issues; $g_{t}^{\text{esc}} \in \{\mathtt{stay\_local}, \mathtt{escalate}\}$ is the binary gating decision for coordination escalation; and $a_{t}$ is the final executed action, encompassing local recovery, local skip, or strictly protocolized public coordination commands.

\subsection{Partitioned Information Architecture}
Our fundamental observation is that the information required to sustain local execution structurally diverges from the information needed to facilitate public collaboration. Mixing them within a unified context pollutes local execution with free-form reflections, conversational history, and redundant coordination signals \citep{chen2025optima,liu2024lost,zhang2023building,xu2026mitigating}. Therefore, we explicitly decouple the execution context into two isolated layers: a Private Execution State and a Public Coordination State.

\subsubsection{\textbf{Private Execution State}}
The private execution state functions as a compact working memory dedicated strictly to local execution and self-healing. To maintain a bounded context length and circumvent state pollution caused by free-form LLM hallucinations \citep{liu2024lost,zhang2023building,xu2026mitigating,liu2025survey}, this state does not append continuous reasoning traces. Instead, it is updated exclusively via system-verified action outcomes triggered at critical events, such as initialization, action completion, recovery mode entry, and mode-level resets. We formalize the private execution state as a tuple:
\begin{equation}
\label{EQ2}
m_t^{\text{priv}} = \langle I_t, T_t, P_t, B_t, H_t \rangle
\end{equation}
where $I_t$ represents the current inventory state; $T_t$ denotes the active sub-task and unfulfilled blueprint requirements; $P_t$ indicates current coordinates and designated work zones; $B_t$ encapsulates the current blockage state (including issue type, missing items, and local recovery candidates); and $H_t$ summarizes the short-term, system-verified action history (e.g., successfully placed, collected, or explicitly failed). This structured representation ensures that the private layer consistently serves immediate local execution without relying on verbose natural language memory.

\subsubsection{\textbf{Public Coordination State}}
The public coordination state is strictly reserved for carrying state-changing collaboration signals. Crucially, the channel between the private and public layers defaults to closed. The system only extracts necessary information from the private state to open a short-term public coordination window when the subsequent gating mechanism deems the collaboration beneficial. When opened, communication within this layer is strictly protocolized to control token consumption and prevent context drift \citep{chen2025optima,zhang2024cut,zhao2024hierarchical,white2025collaborating,yue2025masrouter}. It prohibits casual chatter, generalized planning discussions, and empty ready-checks. Instead, every public message must algorithmically map to a potential task state transition, recording only:
\begin{itemize}[leftmargin=*]
    \item \textbf{State-changing Commands:} Explicit material requests, coordinate references, structural dependencies, and transfer commands.
    \item \textbf{Definitive Replies:} Standardized, explicit responses such as $\mathtt{\textit{CANNOT\_SUPPLY}}$.
\end{itemize}

By abandoning free-form unstructured natural chat in favor of a formalized protocol, we eliminate invalid greetings and redundant confirmations. Once a request is resolved, the short-term window times out, or coordination benefits are exhausted, the public coordination zone immediately closes, cleanly severing the interaction.

\subsection{Gated Collaborative Escalation Policy}
\label{Escalation Policy}
At the core of our methodology lies a three-tiered gating mechanism: \textbf{Heuristic Rules $\rightarrow$ Cost-Sensitive Scoring $\rightarrow$ Gray-Zone LLM Adjudicator.} Its objective extends beyond merely detecting anomalies; it evaluates whether given local issue \textit{warrants} escalation from the private execution layer to a public coordination.

We employ a hierarchical gate rather than a single rule or LLM because collaboration escalation encompasses both structural and semantic dimensions \citep{chen2025optima,zhang2024proagent,chen2024llmarena,li2025parallelized,yue2025masrouter}. A vast majority of cases exhibit deterministic characteristics (e.g., "locally recoverable immediately" or "critical prerequisite blocked") that can be stably handled without LLM intervention. Conversely, boundary cases exist where the cost-benefit tradeoff between local self-healing and public collaboration is too nuanced for simple rules. The three-tiered architecture elegantly decouples these scenarios: rules absorb obvious cases, the scoring function compresses most separable samples, and only genuinely ambiguous boundary cases are delegated to a bounded LLM. This preserves the stability of deterministic control while introducing limited semantic adjudication when necessary.

\subsubsection{\textbf{Issue Detection}}
The gating mechanism is not evaluated at every time step. It is triggered exclusively upon the detection of predefined structured construction issues, denoted as $d_t^{\text{issue}}$. The set of valid structured blockages is defined as:
\begin{equation}
\begin{aligned}
\mathcal{U} = \{ & \mathtt{\textit{missing\_material}}, \mathtt{\textit{support\_failure}}, \\
                 & \mathtt{\textit{dependency\_block}}, \mathtt{\textit{transfer\_needed}}, \\
                 & \mathtt{\textit{co\_craft\_required}} \}
\end{aligned}
\end{equation}
The system initiates the escalation evaluation solely when $d_t^{\text{issue}} \in \mathcal{U}$. This issue-triggered design serves a dual purpose. First, it eliminates the computational overhead and context perturbation of continuously evaluating the "need to collaborate" during smooth execution phases. Second, it anchors the escalation decision to explicit, interpretable, and analyzable structured problems, rather than a generalized, ambiguous state of "agent uncertainty" \citep{dong2024villageragent,chen2024llmarena,liu2025survey}.

\subsubsection{\textbf{Cost-Sensitive Escalation Score}}
\label{score}
For scenarios bypassing intuitive heuristic hard rules, we introduce a cost-sensitive escalation score $S$ to quantify the net benefit of escalating to public:
\begin{equation}
S = w_C C + w_R R + w_I I - w_L L - w_H H
\end{equation}
To ensure robustness against the inherent noise of open-world state estimation, we eschew overly granular continuous variables. Instead, each feature is projected onto a low-cardinality ordinal scale $\in \{0, 1, 2, 3\}$ (see the appendix for details on the mapping of the feature space). This design seamlessly distills heterogeneous environmental signals into a unified, interpretable decision space. Crucially, the scoring function balances two opposing forces.

The positive terms capture the urgency and strategic value of collaboration: \textbf{Criticality ($C$)} encodes the topological importance of the blocked node, prioritizing structural fulcrums on the critical path; \textbf{Coordination Advantage ($R$)} evaluates the comparative efficiency of external help (e.g., a nearby teammate with inventory surplus) over a lengthy local crafting chain; and \textbf{Downstream Impact ($I$)} anticipates the cascading disruptions to the overarching task graph if the anomaly persists. Conversely, the negative terms act as strict regularizers to insulate the system from over-communication. \textbf{Local Recoverability ($L$)} heavily penalizes escalation when deterministic, low-cost local workarounds (e.g., proximate resource gathering or smelting) are viable. Furthermore, the \textbf{Coordination History Penalty ($H$)} serves as a temporal decay factor, explicitly penalizing repeated escalation requests for the same unresolved issue, thereby decisively preventing redundant messaging and coordination deadlocks.

To encode a hierarchical decision priority, we constrain the weights as follows:
\begin{equation}
\label{unequal}
w_{C} > w_{R} \approx w_{I} \approx w_{L} > w_{H}
\end{equation}
This asymmetric weighting dictates that critical path blockages ($C$) dominate the decision baseline, the cost-benefit variables ($R, I, L$) balance operational tradeoffs, and the history penalty ($H$) acts strictly as a corrective regularizer.

\begin{figure}[t]
    \centering
    \includegraphics[width=\linewidth,trim=0  6.5cm 0 8cm,clip]{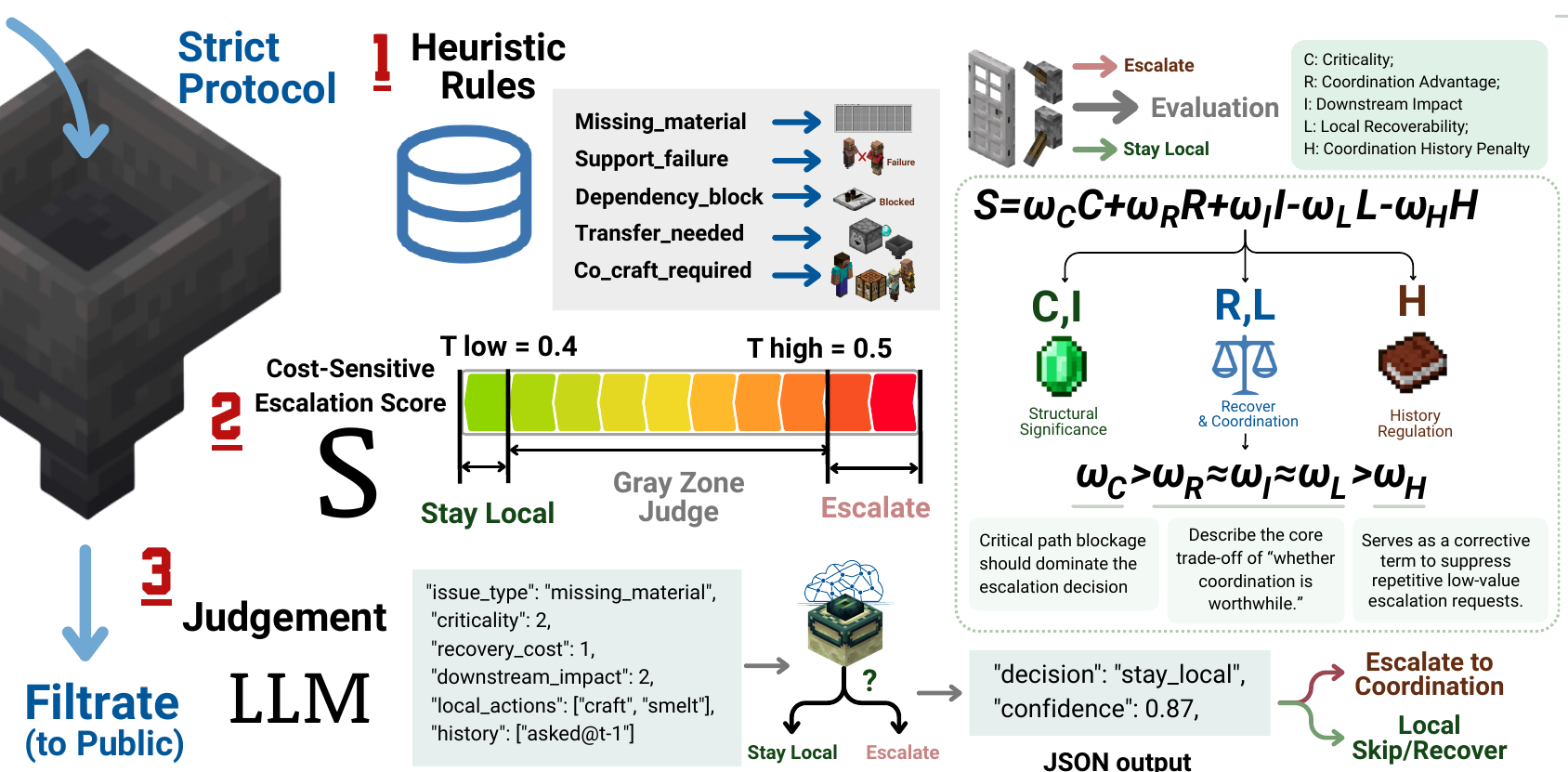}
    \vspace{-2mm}
    \caption{Overview of the Gated Collaborative Escalation Policy}
    \label{fig:my_image}
    \vspace{-2em} 
\end{figure}

\subsubsection{\textbf{Asymmetric Operating Point}}
Upon computing the score $S$, the system employs an asymmetric operating point defined by thresholds $T_{\text{low}}$ and $T_{\text{high}}$:
\begin{equation}
g_t^{\text{esc}} = \begin{cases} 
\text{\textit{stay\_local}}, & \text{if } S_t \le T_{\text{low}} \\ 
\text{\textit{escalate}}, & \text{if } S_t \ge T_{\text{high}} \\ 
\text{\textit{LLM\_Judge}}(\mathbf{f}_t), & \text{if } T_{\text{low}} < S_t < T_{\text{high}} 
\end{cases}
\label{eq:gating_function}
\end{equation}
This asymmetric bias is highly cost-sensitive. In multi-agent systems, the systemic disruption caused by unnecessary collaboration inherently outweighs the cost of conservatively attempting a few additional steps of local recovery \citep{chen2025optima,zhang2024cut,chen2024scalable,li2025parallelized,yue2025masrouter}. Consequently, the system defaults to $\mathtt{stay\_local}$, escalating automatically only when the evidence for collaboration is overwhelmingly strong.

\subsubsection{\textbf{Bounded Gray-Zone LLM Adjudicator}}
\label{LLM Adjudicator}
For ambiguous cases falling into the gray zone ($T_{\text{low}} < S < T_{\text{high}}$), we invoke a bounded LLM adjudicator. Crucially, this model acts neither as a global planner nor an execution controller, but strictly as a binary classifier: it determines whether the issue should remain in the local execution layer or be escalated.

To enforce this boundary, we impose two constraints. On the input side, it receives a strictly structured JSON "decision card" containing feature values and candidate actions, deliberately omitting the main agent's full conversational context to prevent context drift \citep{zhang2024cut,liu2024lost,zhang2023building,xu2026mitigating}. On the output side, it is constrained to return a structured JSON response containing only the decision ($\mathtt{\textit{stay\_local}}$ or $\mathtt{\textit{escalate}}$) and a confidence score \citep{zhang2024proagent,chen2024llmarena}. This meticulously closes the logic loop: obvious cases are absorbed by rules, separable cases by the scoring function, and boundary cases by minimized semantic LLM intervention \citep{chen2025optima,zhang2024cut,zhang2024proagent}. An analysis of its sensitivity to the underlying LLM and call frequency will be presented in appendix.

\subsection{From Decision to Action}
Following the gating decision, control is routed to either the public coordination layer or the local solver. The objective is to ensure that collaboration, when it occurs, is highly efficient, and local handling possesses robust recovery capabilities.

\subsubsection{\textbf{Protocolized Public Coordination}}
If the decision is \textit{\textbf{escalate}}, the system temporarily opens a short-window public coordination channel. This channel exclusively permits strictly formatted protocol commands (e.g., $\mathtt{\textit{REQUEST\_MATERIAL}}$, $\mathtt{\textit{OFFER\_TRANSFER}}$). By eschewing natural language chat for formalized protocols, we eliminate invalid greetings and empty confirmations, ensuring every public message maps directly to algorithmic state transitions \citep{chen2023agentverse,zhang2024cut,zhao2024hierarchical,white2025collaborating,li2025parallelized}. The window closes immediately upon transaction resolution, timeout, or exhaustion of collaboration benefits.

\subsubsection{\textbf{Deterministic Local Solver}}
If the decision is $\mathtt{\textit{\textbf{stay\_local}}}$, control is routed to the deterministic local solver, which provides two operations:
\begin{itemize}[leftmargin=*]
    \item \textbf{$\mathtt{\textbf{\textit{LOCAL\_RECOVER}}}$:} If a clear local recovery chain exists, the system executes an ordered fallback sequence: \textit{$\mathtt{\textbf{craft}} \rightarrow \mathtt{\textbf{smelt}} \rightarrow \mathtt{\textbf{collect}} \rightarrow \mathtt{\textbf{plan}}$.} This embodies a low-to-high cost and near-to-far proximity recovery logic.
    \item \textbf{$\mathtt{\textbf{\textit{LOCAL\_SKIP}}}$:} If the blocked node is non-critical and independent sub-tasks remain viable, the agent temporarily bypasses the blockage to maintain execution momentum.
\end{itemize}

\subsubsection{\textbf{Fallback and Cooldown}}
A defining robustness feature of our framework is that collaboration failures never force system into deadlock. If the public coordination phase yields a $\mathtt{\textit{CANNOT\_SUPPLY}}$ response, times out, or fails to produce valid state changes, the system explicitly updates the history penalty term $H_t$, entering a mandatory communication cooldown for that specific issue type, and immediately routes control back to the local solver.

\subsection{Implementation Details and Parameter Calibration}
\label{Parameter}
To operationalize the gating mechanism without introducing significant computational overhead, the feature extraction process is entirely decoupled from the LLM. We employ lightweight, deterministic modules to project the high-dimensional private execution state $m_t^{\text{priv}}$, the task graph topology, and the team's resource distribution into the low-cardinality ordinal feature space required for the escalation score (as defined in Section~\ref{score}).

Consequently, the parameter space governing the gating logic---comprising the feature weights $\mathbf{w}$ and the asymmetric decision boundaries $(T_{\text{low}}, T_{\text{high}})$---requires rigorous calibration. Rather than relying on heuristic tuning, we formulate this calibration as an offline, cost-sensitive optimization problem over a collected dataset $\mathcal{D}$ of historical execution trajectories \citep{chen2025optima,chen2024scalable,chen2024llmarena,yue2025masrouter}.

To identify a Pareto-optimal configuration $\Theta^* = \{\mathbf{w}^*, T_{\text{low}}^*, T_{\text{high}}^*\}$ that balances local autonomy with collaborative efficiency, the objective is designed to maximize the Task Success Rate (TSR) subject to strict penalty constraints on system overhead \citep{chen2025optima,chen2024scalable,li2025parallelized}. Formally, the calibration is solved via grid search to maximize the following expected utility:
\begin{equation}
\begin{split}
\Theta^* = \arg\max_{\Theta} \mathbb{E}_{\tau \sim \mathcal{D}} \Big[ & \text{TSR}(\tau) - \lambda_1 C_{\text{time}}(\tau) \\
& - \lambda_2 C_{\text{redundant}}(\tau) - \lambda_3 C_{\text{LLM}}(\tau) \Big]
\end{split}
\label{eq:cost_optimization}
\end{equation}
where for a given trajectory $\tau$, $C_{\text{time}}$ denotes the average local recovery time, $C_{\text{redundant}}$ represents the frequency of zero-yield escalations, and $C_{\text{LLM}}$ measures the token consumption of the LLM adjudicator. The penalty coefficients $\lambda_i$ explicitly enforce the conservative bias of our architecture by penalizing redundant interactions. The exact parameter values derived from this optimization, as well as an analysis of parameter generalization across different domains, are detailed in the Appendix.
\section{Experiments}
\label{sec:experiments}

To systematically evaluate the effectiveness of our proposed Partitioned Information Architecture and Gated Escalation Mechanism, we design our experiments to answer the following core research questions (RQs):
\begin{itemize}[leftmargin=*]
    \item \textbf{RQ1 (Overall Performance):} Can our framework maintain robust task success rates while reducing execution redundancy in both standard and highly collaborative environments?
    \item \textbf{RQ2 (Coordination Efficiency):} How effectively does the gating mechanism filter out unnecessary communication and improve local self-recovery capabilities?
    \item \textbf{RQ3 (Ablation):} What are the individual contributions of the partitioned architecture and the multi-tiered gating strategies to the overall system performance?
\end{itemize}

\subsection{Datasets and Baselines}
\label{subsubsec:datasets}
\subsubsection{\textbf{Datasets}}
To systematically evaluate our framework, we conduct experiments on two representative multi-agent Minecraft platforms: MindCraft and VillagerBench \citep{white2025collaborating,dong2024villageragent}. We categorize our evaluation into two distinct settings to disentangle foundational execution from complex coordination.

\textbf{Standard Benchmarks} (Weak Collaboration): As a control group, we utilize the standard built-in benchmarks of both platforms, comprising 7 native long-horizon tasks in MindCraft and a dynamic architectural blueprint library in VillagerBench. Evaluated via their respective auto-scoring pipelines, these settings provision agents with sufficient initial resources. 

\textbf{Custom Datasets} (High Collaboration): To rigorously stress-test coordination dynamics, we construct custom datasets that inject explicit material splits and dependency bottlenecks, forcing agents to navigate communication tradeoffs. \textbf{For MindCraft}, we curate 200 episodes evaluated via custom scripts, evenly divided into four distinct stress-test scenarios (as shown in Table \ref{tab:dataset_stats}): \textbf{(A)} \textit{Locally Recoverable}, \textbf{(B)} \textit{Necessary Collaboration}, \textbf{(C)} \textit{Gray-zone Decision}, and \textbf{(D)} \textit{Collaboration Failure Robustness}. \textbf{For VillagerAgent}, we construct a custom high-collaboration construction split on top of the native benchmark, preserving the original blueprint-family and complexity distribution. Subsequently, we ensured that success depended entirely on effective cross-agent communication by strictly allocating key resources to isolated boxes or inventories. We disabled the default teammate state sharing to increase coordination pressure, while keeping the underlying construction objectives unchanged. For specific definitions and examples of material partitioning and dependency bottlenecks, please refer to the Appendix.

\begin{table}[htbp]
    \centering
    \caption{MindCraft Custom Dataset Statistics}
    \label{tab:dataset_stats}
    \begin{tabular}{@{}ll@{}}
        \toprule
        \textbf{Metric} & \textbf{Value} \\
        \midrule
        templates & 40 \\
        seeds\_per\_template & 5 \\
        total\_episodes & 200 \\
        two\_agent\_episodes & 120 \\
        three\_agent\_episodes & 80 \\
        class\_A\_episodes & 50 \\
        class\_B\_episodes & 50 \\
        class\_C\_episodes & 50 \\
        class\_D\_episodes & 50 \\
        \bottomrule
    \end{tabular}
\vspace{-1.5em} 
\end{table}

\subsubsection{\textbf{Baselines}}
We benchmark our approach against two representative multi-agent LLM paradigms:
\begin{itemize}[leftmargin=*]
    \item \textbf{MindCraft (FlatComm):} Employs full free-form communication, where agents broadcast issues immediately upon encountering uncertainty.
    \item \textbf{VillagerAgent (DAG):} Utilizes a centralized Directed Acyclic Graph (DAG) planner to allocate tasks, representing rigid, top-down coordination.
\end{itemize}

\subsection{Evaluation Metrics}

\subsubsection{\textbf{Macro-Performance Indicators}}
We adopt Task Success Rate (TSR) and Completion Steps (CS) as our macro-performance indicators to quantify overall system efficacy.

\textbf{Task Success Rate (TSR).} TSR measures the final construction quality of the generated structure. It reflects how accurately the system completes the target blueprint at the block level, and serves as the primary indicator of task success.

\textbf{Completion Steps (CS).} CS measures the overall execution cost of solving a task. It counts the total number of environment-interacting actions taken by all agents, which is used to assess whether the proposed selective collaboration mechanism reduces unnecessary action overhead.

\subsubsection{\textbf{Mechanistic Metrics for Coordination Dynamics}}

Beyond final performance, we further introduce several mechanistic metrics to analyze how the system decides between local handling and public coordination, and whether the proposed gated collaboration strategy truly improves coordination quality.

\textbf{Local Resolution Rate (LRR).} LRR measures the proportion of resolved issues that are handled entirely within the private execution loop. It reflects whether the framework genuinely prioritizes local recovery before requesting help.

\textbf{Unnecessary Escalation Rate (UER).} UER measures how often the system triggers public coordination when a feasible local solution was actually available at decision time. A lower UER indicates that collaboration is invoked more selectively and only when truly needed.

\textbf{Effective Communication Rate (ECR).} ECR measures the proportion of coordination episodes that produce a verified, task-relevant state change within a short time window. A higher ECR indicates that communication is not only sparse, but also effective.

\textbf{Recovery Success Rate (RSR).} RSR measures the proportion of encountered issues that are eventually resolved once any recovery mechanism has been activated. It captures the overall robustness of the recovery pipeline.

\subsection{Implementation Details}
We evaluate two representative multi-agent Minecraft baselines, \textbf{MindCraft/MineCollab} and \textbf{VillagerAgent}, under their official environments while keeping the control stack conceptually aligned. Both rely on \textbf{Mineflayer} and the \textbf{Prismarine} ecosystem for low-level motor control, enabling agents to execute dynamically generated free-form actions through a unified action interface. They adhere to Minecraft versions 1.21.1 / 1.19.2, respectively, to meet the requirements of their native backends and built-in evaluators. Across modules, we employ a single backbone MLLM model based on the original paper’s setup to handle dialogue, planning, and execution-related reasoning (the former uses \textbf{GPT-4o}; the latter uses \textbf{GPT-4-1106-preview}), while retrieval is performed via document and contextual example lookups. 

To ensure a fair comparison, all hybrid models were rigorously reimplemented within the corresponding frameworks. Compared to the baseline, they share exactly the same action interfaces, prompt budgets, and environmental settings, thereby ensuring that any performance differences are entirely attributable to the underlying coordination strategies.

\begin{table}[t]
\centering
\caption{Overall Performance on Standard and Custom Construction Tasks. TSR: Task Success Rate (\%); CS: Completion Steps.}
\label{tab:overall_performance}
\renewcommand{\arraystretch}{1.15} 
\resizebox{\columnwidth}{!}{
\begin{tabular}{l | cc | cc}
\toprule
\multirow{2}{*}{Method} & \multicolumn{2}{c|}{\textbf{Standard Setting}} & \multicolumn{2}{c}{\textbf{Custom Setting}} \\
\cmidrule(lr){2-3} \cmidrule(lr){4-5} 
 & TSR $\uparrow$ & CS $\downarrow$ & TSR $\uparrow$ & CS $\downarrow$ \\
\midrule
\multicolumn{5}{c}{\textit{Free-form Communication Framework}} \\
\midrule
MindCraft (Baseline)       & 31.1 & 396 & 21.5 & 184 \\
MindCraft + \textbf{Ours}  & \textbf{35.7} & \textbf{294} & \textbf{32.8} & \textbf{134} \\
\midrule
\multicolumn{5}{c}{\textit{Structured Planning Framework}} \\
\midrule
VillagerAgent (Baseline)   & 36.45 & 103 & 22.04 & 145 \\
VillagerAgent + \textbf{Ours}& \textbf{42.76} & \textbf{76} & \textbf{34.56} & \textbf{92} \\
\bottomrule
\end{tabular}
}

\end{table}

\subsection{Main Results (RQ1)}
Table \ref{tab:overall_performance} reports the comparative performance of our framework and the baselines across both Standard and Custom construction tasks.

\begin{table}[t]
\centering
\small 
\renewcommand{\arraystretch}{1.05} 
\setlength{\tabcolsep}{4.5pt} 
\caption{In-depth Analysis of Coordination Efficiency and Local Autonomy. LRR: Local Resolution Rate; UER: Unnecessary Escalation Rate; ECR: Effective Communication Rate; RSR: Recovery Success Rate. All metrics are in \%.}
\label{tab:coordination_efficiency}
\begin{tabular}{@{}ll cccc@{}}
\toprule
\textbf{Setting} & \textbf{Method} & LRR $\uparrow$ & UER $\downarrow$ & ECR $\uparrow$ & RSR $\uparrow$ \\
\midrule
\multicolumn{6}{@{}l}{\textit{Free-form Communication Framework}} \\
\midrule 
\multirow{2}{*}{Standard} 
& MindCraft       & 52.3 & 67.8 & 38.9 & 58.7 \\
& \textbf{+ Ours} & \textbf{89.7} & \textbf{11.3} & \textbf{83.9} & \textbf{84.8} \\
\cmidrule(l){2-6} 
\multirow{2}{*}{Custom}   
& MindCraft       & 43.5 & 61.3 & 46.2 & 50.2 \\
& \textbf{+ Ours} & \textbf{76.5} & \textbf{16.0} & \textbf{81.3} & \textbf{78.1} \\
\midrule
\multicolumn{6}{@{}l}{\textit{Structured Planning Framework}} \\
\midrule
\multirow{2}{*}{Standard} 
& VillagerAgent   & 41.8 & 78.5 & 41.2 & 52.1 \\
& \textbf{+ Ours} & \textbf{86.5} & \textbf{14.2} & \textbf{84.5} & \textbf{83.2} \\
\cmidrule(l){2-6}
\multirow{2}{*}{Custom}   
& VillagerAgent   & 35.4 & 73.3 & 35.8 & 40.5 \\
& \textbf{+ Ours} & \textbf{73.2} & \textbf{17.4} & \textbf{78.2} & \textbf{76.5} \\
\bottomrule
\end{tabular}

\end{table}

\textbf{Robustness in Resource-Constrained Environments}. As shown in Table \ref{tab:overall_performance}, existing baselines achieve competitive Task Success Rates (TSR) in the Standard setting. However, they experience a severe performance drop when transitioned to the Custom setting. For example, the TSR of VillagerAgent falls sharply from \textbf{36.45\%} to \textbf{22.04\%}. This degradation occurs because standard baselines struggle to manage hidden states and distributed resources, frequently leading to coordination deadlocks \citep{dong2024villageragent,sun2025collab}. Conversely, our framework remains highly robust. By integrating our proposed mechanisms, it sustains a strong TSR of \textbf{34.56\%} under Custom conditions. This highlights our method's adaptability to environments that strictly require explicit coordination.

\textbf{Execution Efficiency}. Beyond improving success rates, our approach significantly optimizes execution efficiency. Across all tasks, our method consistently requires fewer Completion Steps (CS) than the unmodified baselines. Notably, in the Custom setting, it reduces the CS of VillagerAgent from \textbf{145} to \textbf{92}. This improvement validates our core hypothesis. By restricting premature and unstructured communication, our agents maintain steady execution momentum. As a result, they effectively bypass the cascading delays typically caused by global state synchronization.

\subsection{In-depth Analysis of Efficiency (RQ2)}
To understand the efficiency gains of our framework, we analyze the agent communication behaviors detailed in Table \ref{tab:coordination_efficiency}.

\textbf{Enhancing Local Autonomy}. Baseline models typically default to global communication when encountering errors. This over-reliance results in a low Local Resolution Rate (LRR) of approximately \textbf{40\%} and a high Unnecessary Escalation Rate (\textbf{UER > 70\%}). In contrast, our framework explicitly separates private execution states from public coordination. This decoupling allows agents to fully utilize the deterministic local solver. As a result, the LRR increases significantly, ranging from \textbf{30\% to 50\%}. This demonstrates that the local execution loop successfully absorbs the majority of minor anomalies (e.g., simple missing materials) without requiring global intervention.

\textbf{Maximizing Collaboration Utility}. When communication is necessary, our framework ensures it is highly purposeful. The proposed Gated Escalation Mechanism drives the Effective Communication Rate (ECR) by \textbf{40\% to 50\%}, substantially outperforming the baselines. Furthermore, the Recovery Success Rate (RSR) shows a notable improvement. These results indicate that our cost-sensitive, protocol-based public channel successfully filters out conversational noise. Consequently, broadcast messages are not wasted; instead, they directly trigger actionable collaborative state transitions and prevent deadlocks. Notably, these gains are accompanied by a reduction in the number of escalation events and messages.

\begin{table}[t]
\caption{Ablation study of different components on the custom setting. TSR: Task Success Rate, CS: Completion Steps, ECR: Effective Communication Ratio, Msg: Total Messages.}
\label{tab:ablation_custom}
\centering
\resizebox{\columnwidth}{!}{%
\begin{tabular}{@{} l c ccc cccc @{}}
\toprule
\multirow{2}{*}{\textbf{Variant}} & \multirow{2}{*}{\shortstack{\textbf{Part.}\\\textbf{Arch.}}} & \multicolumn{3}{c}{\textbf{Gating Tiers}} & \multicolumn{4}{c}{\textbf{Metrics}} \\
\cmidrule(lr){3-5} \cmidrule(l){6-9} 
 & & Rule & Score & LLM & TSR$\uparrow$ & CS$\downarrow$ & ECR$\uparrow$ & Msg$\downarrow$ \\
\midrule
Base (MindCraft)      & & & & & 21.5 & 184 & 46.2 & 78 \\
\midrule
w/o Partition         & & $\checkmark$ & $\checkmark$ & $\checkmark$ & 27.1 & 158 & 72.4 & 42 \\
w/o Gating            & $\checkmark$ & & & & 23.3 & 172 & 51.8 & 66 \\
\midrule
Gate: Rule            & $\checkmark$ & $\checkmark$ & & & 27.5 & 156 & 58.5 & 51 \\
Gate: Rule + Score    & $\checkmark$ & $\checkmark$ & $\checkmark$ & & 31.4 & 141 & 78.3 & 14 \\
\midrule
\textbf{Full Model}   & $\checkmark$ & $\checkmark$ & $\checkmark$ & $\checkmark$ & \textbf{32.8} & \textbf{134} & \textbf{81.3} & \textbf{11} \\
\bottomrule
\end{tabular}%
}

\end{table}

\subsection{Ablation Study (RQ3)}
To isolate the contributions of each proposed component, we perform an ablation study on the Custom Datasets using MindCraft as the baseline. Table \ref{tab:ablation_custom} summarizes these results.

\textbf{Impact of Partitioned Architecture}. We first remove the partitioned architecture ("w/o Partition"). This merges private reasoning and public coordination into a shared context window. As a result, the average number of messages (Msgs) surges, and the TSR noticeably drops. This empirically demonstrates that context pollution and hallucination are severe bottlenecks in standard, unpartitioned multi-agent MLLMs \citep{xu2026mitigating,liu2025survey}.

\begin{figure*}[t]
    \centering
    \includegraphics[width=0.88\linewidth,trim=0 0.8cm 0 0,clip]{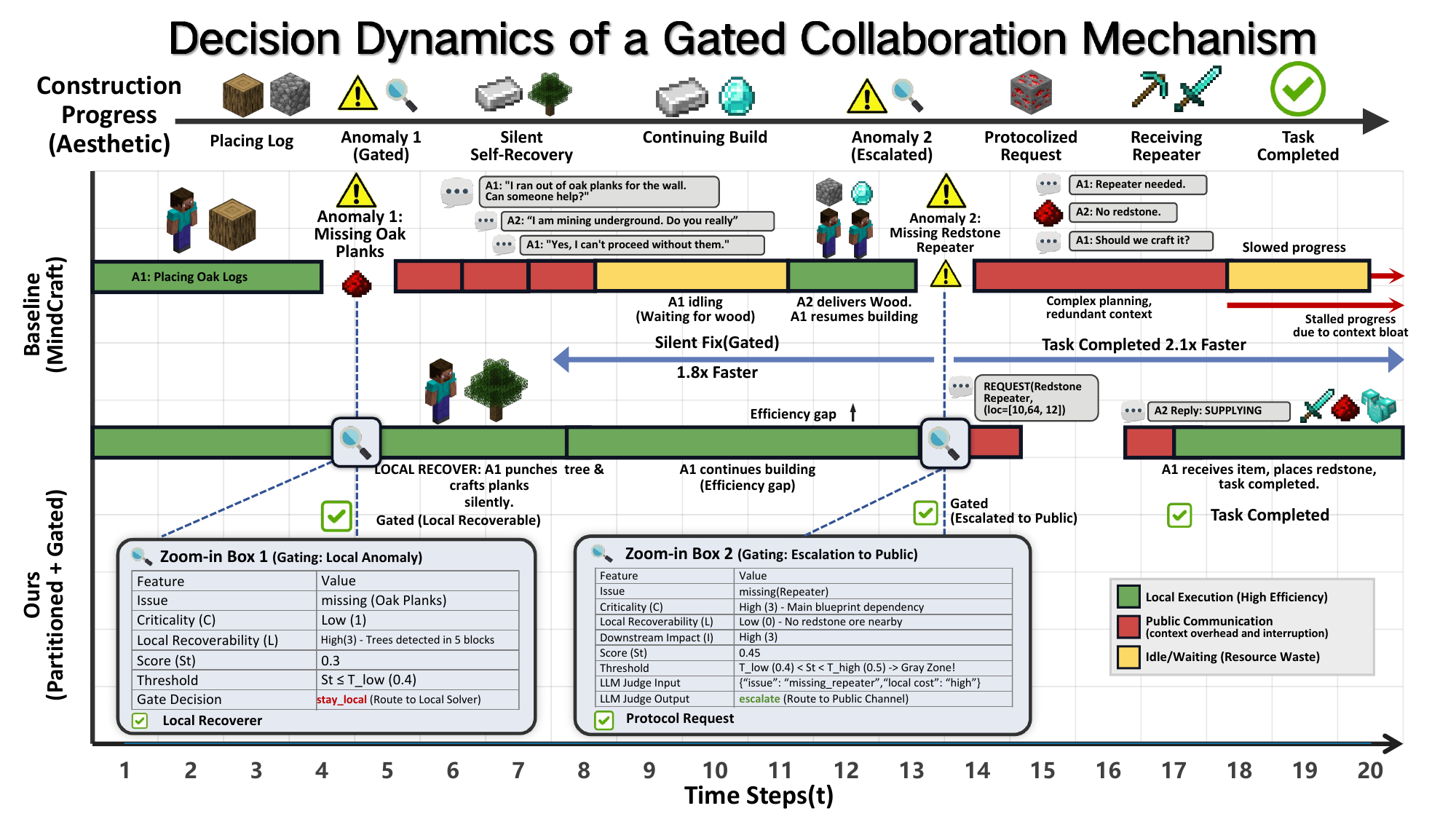}
    \caption{Qualitative comparison of decision dynamics between the Baseline (Mindcraft) and our Gated Escalation framework. (Top) The baseline defaults to free-form communication. (Bottom) Our model utilizes a 3-tier gate.}
    \label{fig:my_image}
 
\end{figure*}

\textbf{Effectiveness of Multi-tiered Gating}. We also evaluate our gating mechanism by incrementally adding its components. This progressive setup reveals a steady trajectory of performance improvements:
\begin{itemize}[leftmargin=*]
    \item \textbf{Rules Only}: Relying solely on heuristic rules successfully filters out basic errors. However, it struggles to evaluate complex tradeoffs, leading to a suboptimal ECR and TSR.
    \item \textbf{Rules + Score}: Adding the cost-sensitive scoring layer yields the largest gain: communication is sharply reduced (Msgs from \textbf{51 to 14}), while both TSR and ECR improve substantially. 
    \item \textbf{Full Model}: Finally, incorporating the bounded MLLM adjudicator resolves ambiguous edge cases. This complete configuration achieves the optimal balance: the highest TSR (\textbf{32.8\%}) alongside minimal communication overhead, showing that selective collaboration is most effective when lightweight rules, quantitative scoring, and bounded semantic judgment are combined.
\end{itemize}

\subsection{Case Study: Decision Dynamics in Action}
To qualitatively illustrate our framework's decision dynamics, Fig.~\ref{fig:my_image} contrasts our Gated Escalation Mechanism with the free-form baseline (MindCraft) in a resource-constrained construction scenario.

\textbf{\textit{Case 1: Silent Fix via Local Recovery.}} At $t=5$, Agent 1 encounters a minor anomaly (missing oak planks). While the baseline broadcasts indiscriminately—interrupting its teammate (red blocks), our module ($d_{t}^{\text{issue}}$) intercepts the issue. Given high local recoverability ($L$) and low criticality ($C$), the computed score falls below the conservative threshold ($S \le T_{\text{low}}$). This deterministically triggers stay\_local, allowing Agent 1 to silently craft the planks (\textit{LOCAL\_RECOVER}). This perfectly preserves the public coordination state and both agents' execution momentum.

\textbf{\textit{Case 2: Cost-Sensitive Escalation.}} At $t=14$, a critical dependency blockage (missing repeater) stalls the baseline, which degenerates into unstructured planning and prolonged idling (yellow blocks). Conversely, our framework extracts structured evidence: high criticality ($C=3$), high downstream impact ($I=3$), and zero local recoverability ($L=0$), shifting the score into the gray zone ($T_{\text{low}} < S < T_{\text{high}}$). The bounded LLM adjudicator evaluates this and authorizes an escalate action. By enforcing a protocolized request (\textit{REQUEST(Redstone\_Repeater)}), our method strictly eliminates conversational noise, drastically reducing Completion Steps (CS) and maximizing the Effective Communication Rate (ECR).

\textbf{\textit{Case 3: Robust Fallback.}} When an escalated request yields a \textit{CANNOT\_SUPPLY} response, the baseline frequently deadlocks into repeated querying or permanent stalling. In stark contrast, our framework leverages this explicit failure signal to update the history penalty ($H$) and initiate a communication cooldown. Control deterministically reverts to the local solver (\textit{LOCAL\_SKIP}), enabling the agent to bypass the blocked node and construct an orthogonal blueprint subregion. This explicit fallback prevents cascading delays and ensures robust fault tolerance.

Together, these findings clearly validate the necessity of our hierarchical gating design.
\section{Conclusion}
\label{sec:conclusion}
In this work, we challenge the prevalent "communication-first" paradigm in MLLM-driven multi-agent systems by introducing a Partitioned Information Architecture and a Gated Escalation Mechanism. Rather than treating local anomalies as automatic triggers for global interaction, our framework explicitly decouples private execution memory from public coordination states, empowering agents to autonomously weigh the feasibility of local self-recovery against the overhead of collaboration. Extensive experiments in long-horizon Minecraft tasks demonstrate that by transforming communication from a default reflex into a selective, calculated decision, our system significantly outperforms existing baselines, delivering not only higher-quality blueprint completion and shorter execution chains but also demonstrating superior communication quality and robustness. Ultimately, our findings establish a critical insight for embodied AI: scalable collaborative efficiency stems not from maximizing communication volume, but from the rigorous, dynamic governance of interaction boundaries.

\bibliographystyle{ACM-Reference-Format}
\bibliography{sample-base}

\clearpage 
\appendix  

\section{Appendix}

\subsection{Detailed Algorithmic Execution Flow}
\label{app:execution_flow}

Algorithm \ref{alg:gated_escalation} formalizes the complete execution loop of a single agent within our partitioned information architecture. The core innovation lies in the three-tiered Gated Collaborative Escalation Policy (Section \ref{Escalation Policy}), which evaluates whether a local anomaly warrants global disruption.

\begin{algorithm}[htbp]
\caption{Execution Loop with Gated Collaborative Escalation}
\label{alg:gated_escalation}
\begin{algorithmic}[1]
\Require Agent $a$, observation $o_t$, private state $m_{t-1}^{\text{priv}}$, thresholds $T_{\text{low}}, T_{\text{high}}$
\Ensure Executed action $a_t$, updated private state $m_t^{\text{priv}}$

\Statex \textbf{\textit{Phase 1: Private State Update \& Issue Detection}}
\State $m_t^{\text{priv}} \gets \textsc{UpdatePrivateState}(o_t, m_{t-1}^{\text{priv}})$
\State $d_t^{\text{issue}} \gets \textsc{DetectIssue}(m_t^{\text{priv}})$
\If{$d_t^{\text{issue}} \notin \mathcal{U}$} \Comment{No structural anomaly detected}
    \State \Return $(\textsc{StandardExecution}(m_t^{\text{priv}}), m_t^{\text{priv}})$
\EndIf

\Statex \textbf{\textit{Phase 2: Cost-Sensitive Scoring (Tier 1 \& 2)}}
\State $(C, R, I, L, H) \gets \textsc{ExtractFeatures}(m_t^{\text{priv}}, \text{TaskGraph})$
\State $S_t \gets w_C C + w_R R + w_I I - w_L L - w_H H$

\Statex \textbf{\textit{Phase 3: Asymmetric Decision \& Adjudication (Tier 3)}}
\If{$S_t \le T_{\text{low}}$}
    \State $g_t^{\text{esc}} \gets \texttt{stay\_local}$ \Comment{Absorbable by local solver}
\ElsIf{$S_t \ge T_{\text{high}}$}
    \State $g_t^{\text{esc}} \gets \texttt{escalate}$ \Comment{High collaborative yield}
\Else
    \State $g_t^{\text{esc}} \gets \textsc{BoundedLLMJudge}(S_t, d_t^{\text{issue}})$ \Comment{Gray-zone}
\EndIf

\Statex \textbf{\textit{Phase 4: Action Routing \& Graceful Fallback}}
\If{$g_t^{\text{esc}} = \texttt{escalate}$}
    \State $a_t \gets \textsc{ProtocolizedCoordination}(d_t^{\text{issue}})$
    \State $res \gets \textsc{Execute}(a_t)$
    \If{$res \in \{\texttt{TIMEOUT}, \texttt{CANNOT\_SUPPLY}\}$}
        \State $H_t \gets H_t + \lambda$ \Comment{Add history penalty for cooldown}
        \State $a_t \gets \textsc{LocalSolver}(d_t^{\text{issue}}, m_t^{\text{priv}})$ \Comment{Fallback}
    \EndIf
\Else
    \State $a_t \gets \textsc{LocalSolver}(d_t^{\text{issue}}, m_t^{\text{priv}})$
\EndIf

\State \Return $(a_t, m_t^{\text{priv}})$
\end{algorithmic}
\end{algorithm}

\textbf{Mechanistic Insights of the Algorithm:}
\begin{itemize}[leftmargin=*, parsep=0pt]
    \item \textbf{Phase 1 (Issue-Triggered Evaluation):} To prevent continuous context perturbation, the escalation logic is only activated when a structural anomaly ($d_t^{\text{issue}} \in \mathcal{U}$) is strictly detected. Otherwise, the agent safely remains in standard execution.
    \item \textbf{Phase 2 \& 3 (Cost-Benefit Adjudication):} The framework prevents distress broadcasting by computing score $S_t$. The asymmetric bounds ($T_{\text{low}}, T_{\text{high}}$) heavily bias the agent towards \textit{stay\_local}. Only high-yield anomalies or ambiguous cases (passed to the bounded LLM) can interrupt the global state.
    \item \textbf{Phase 4 (Deadlock Prevention via Cooldown):} If public coordination fails (\textit{CANNOT\_SUPPLY}), the system applies a penalty ($\lambda$) to $H_t$, enforcing a communication cooldown. It deterministically routes back to the deterministic local solver (e.g., triggering a \textit{LOCAL\_SKIP} ), averting infinite collaborative deadlocks.
\end{itemize}

\newcommand{\suppboxed}[1]{%
    \setlength{\fboxsep}{6pt}%
    \fcolorbox{black}{gray!4}{%
        \begin{minipage}{0.97\linewidth}
        \footnotesize
        #1
        \end{minipage}
    }%
}

\subsection{Deterministic Mapping of the Escalation Feature Space}
\label{Mapping of the Escalation Feature Space}
As introduced in Section \ref{score} and of the main text, the cost-sensitive escalation score $S$ relies on five key features: Criticality (\textbf{$C$}), Coordination Advantage (\textbf{$R$}), Downstream Impact (\textbf{$I$}), Local Recoverability (\textbf{$L$}), and Coordination History Penalty (\textbf{$H$}). To ensure robustness against the inherent noise of open-world state estimation and to strictly bound the computational overhead, we explicitly eschew continuous representations and LLM-based heuristic evaluations for feature extraction. Instead, we introduce lightweight, deterministic algorithmic modules. These modules project the high-dimensional private execution state $m_t^{priv} = \langle I_t, T_t, P_t, B_t, H_t \rangle$ alongside the task graph topology into a unified, low-cardinality ordinal scale $e \in \{0, 1, 2, 3\}$. 

This quantization strategy guarantees that the escalation mechanism maintains mathematical interpretability, avoids the hallucination risks associated with free-form LLM reasoning, and achieves near-zero latency during dynamic execution. The explicit mapping criteria for each feature are systematically defined in Table \ref{tab:feature_mapping}. 

\begin{table*}[htbp]
\centering
\caption{Deterministic Mapping Rules for the Escalation Feature Space. Positive terms ($C, R, I$) drive the system toward public collaboration, while negative penalty terms ($L, H$) strictly regularize the model to favor local autonomy and prevent coordination deadlocks.}
\label{tab:feature_mapping}
\renewcommand{\arraystretch}{1.2}
\small
\begin{tabular}{@{}cclp{10.5cm}@{}}
\toprule
\textbf{Category} & \textbf{Feature} & \textbf{Scale} & \textbf{Deterministic Condition \& Domain Example} \\
\midrule

\multirow{12}{*}{\rotatebox{90}{\textbf{Escalation Drivers (Positive)}}} 
& \multirow{4}{*}{\shortstack{\textbf{Criticality}\\($C$)}} 
& $0$ & \textbf{Non-essential/Aesthetic:} The blocked node has no dependent descendants in the task graph (e.g., placing decorative elements like a flower pot). \\
& & $1$ & \textbf{Leaf Dependency:} The node is a prerequisite for a localized, independent sub-structure with depth $\le 1$ (e.g., a wall block supporting a specific window). \\
& & $2$ & \textbf{Structural Support:} The node serves as a core structural prerequisite for major upcoming blueprint components (e.g., placing a load-bearing pillar). \\
& & $3$ & \textbf{Critical Path Bottleneck:} The node lies on the singular critical path; no parallel sub-tasks can be initiated until resolved (e.g., a foundational redstone trigger). \\
\cmidrule{2-4}

& \multirow{4}{*}{\shortstack{\textbf{Coordination}\\ \textbf{Advantage}\\($R$)}} 
& $0$ & \textbf{No Advantage:} No teammates are within a viable radius ($>50$ blocks), or team broadcasts indicate zero surplus of the required material. \\
& & $1$ & \textbf{Marginal Advantage:} A teammate is nearby but only possesses raw materials, requiring additional crafting steps before transfer. \\
& & $2$ & \textbf{Significant Advantage:} A teammate is within interaction range and holds the exact required material in their inventory surplus. \\
& & $3$ & \textbf{Overwhelming Advantage:} A teammate is in the immediate vicinity ($<10$ blocks) with the item ready to drop, reducing recovery latency to near zero. \\
\cmidrule{2-4}

& \multirow{4}{*}{\shortstack{\textbf{Downstream}\\ \textbf{Impact}\\($I$)}} 
& $0$ & \textbf{Isolated:} The blockage exclusively delays the current agent's immediate local action sequence without affecting the global blueprint. \\
& & $1$ & \textbf{Minor Disruption:} Resolving the issue locally forces the agent to deviate from its optimal path, slightly increasing the global makespan. \\
& & $2$ & \textbf{Moderate Disruption:} The blockage initiates a prolonged local recovery chain, potentially leaving a dependent teammate idling. \\
& & $3$ & \textbf{Global Halt:} The anomaly paralyzes the execution flow of all agents, necessitating immediate cross-agent resolution to resume global progress. \\
\midrule

\multirow{8}{*}{\rotatebox{90}{\textbf{Regularizers}}} 
& \multirow{4}{*}{\shortstack{\textbf{Local}\\ \textbf{Recoverability}\\($L$)}} 
& $0$ & \textbf{Impossible:} The required resource cannot be generated locally (e.g., an agent strictly isolated in a biome devoid of required ores). \\
& & $1$ & \textbf{High Cost:} Local recovery demands a complex execution chain (\textit{craft $\rightarrow$ smelt $\rightarrow$ collect}) or traversing a significant spatial distance. \\
& & $2$ & \textbf{Medium Cost:} The agent possesses raw materials and proximal access to a crafting table, requiring only a brief crafting action to resolve the anomaly. \\
& & $3$ & \textbf{Immediate Resolution:} The required material is already physically present in the immediate environment (e.g., dropped nearby or in an adjacent chest). \\
\cmidrule{2-4}

& \multirow{4}{*}{\shortstack{\textbf{History}\\ \textbf{Penalty}\\($H$)}} 
& $0$ & \textbf{Fresh Issue:} No previous public coordination requests have been broadcast regarding this specific node in the current execution phase. \\
& & $1$ & \textbf{Recent Unanswered Request:} A prior request was escalated but timed out without a definitive reply, triggering a soft communication cooldown. \\
& & $2$ & \textbf{Repeated Escalation:} The agent has queried this issue multiple times with zero coordination yield, heavily dampening the probability of re-escalation. \\
& & $3$ & \textbf{Explicit Rejection:} The agent recently received a definitive \texttt{\textit{CANNOT\_SUPPLY}} signal. The system enforces a mandatory local fallback execution loop. \\
\bottomrule
\end{tabular}
\end{table*}

\subsection{Prompt and State Examples}
\label{sec:supp_prompt_state_examples}

To improve reproducibility without overloading the supplement, we provide representative prompt interfaces and rendered state examples from our implementation. These examples instantiate the two design principles of our method: \emph{bounded decomposition} at the planning layer and \emph{partitioned state exposure} at the execution layer. Specifically, the planner is constrained to emit a fixed JSON-style decomposition, while each agent receives a prompt in which private execution memory and public coordination records are strictly separated. Long recipes, inventories, and environment summaries are truncated for readability.

\subsubsection{\textbf{Task-Level Prompt:} Bounded Decomposition Interface}

For construction tasks, the task manager receives a compact scene summary together with a structured blueprint recipe, and is required to output only a small set of subtasks with fixed keys. This bounded interface prevents unconstrained free-form planning and ensures that the planner produces a decomposition that can be consumed deterministically by downstream agents.

\begin{figure*}[t]
\centering
\suppboxed{
\textbf{System role.} You are the task manager for a multi-agent Minecraft construction task. Decompose the current blueprint objective into a small number of executable subtasks and arrange them in chronological order. Return only a valid JSON array.

\vspace{0.35em}
\textbf{Input context.}

{\ttfamily
Scene summary: Sign info = minecraft/templates/village\_desert\_streets\_turn\_01 \par
High-level objective: collaboratively place blocks according to the blueprint \par
Blueprint recipe: [material=smooth\_sandstone, facing=A, position=[-9,-60,0], ...]
}

\vspace{0.35em}
\textbf{Output contract.}

{\ttfamily
Return ONLY a JSON array with 1--2 subtask objects. \par
Each object must contain exactly: \par
"id", "description", "milestones", "retrieval\_paths", \par
"required\_subtasks", "assigned\_agents".
}
}
\caption{Representative planning-layer prompt. The planner is exposed only to a compact scene summary and a structured blueprint specification, and is forced to return a bounded decomposition interface rather than an unconstrained free-form plan.}
\label{fig:task_prompt}
\end{figure*}

\subsubsection{\textbf{Execution Prompt:} Private Memory vs. Public Board}

At execution time, each agent receives a task-conditioned prompt consisting of four logically distinct inputs: the local task assignment, the private execution state, the environment summary, and the public coordination board. Under our framework, \textit{agent\_state} is rendered exclusively from verified private working memory, whereas \textit{other\_agents} is rendered exclusively from protocol-normalized public records. This design directly implements the partitioned information architecture in the main paper.

\begin{figure*}[t]
\centering
\suppboxed{
\textbf{Agent.} Alice

\vspace{0.35em}
\textbf{Local task data.} {\ttfamily []}

\vspace{0.35em}
\textbf{Public coordination board.} {\ttfamily ["No public coordination records yet."]}

\vspace{0.35em}
\textbf{Private execution state (verified working memory).}

{\ttfamily
I\_t: inventory = dirt*143, ladder*143, oak\_planks*10, ... \par
T\_t: active\_subtask = none; unfinished\_requirements = [] \par
P\_t: current = [-4,-59,1]; work\_region = [-4,-59,1] \par
B\_t: blocker = none; recovery\_mode = none; cooldown = none \par
H\_t: verified\_history = none
}

\vspace{0.35em}
\textbf{Environment summary.} Nearby entities, visible blocks, containers, furnaces, and sign information.

\vspace{0.35em}
\textbf{Coordination policy addendum.} Treat \texttt{agent\_state} as private verified working memory. Treat \texttt{other\_agents} as the public coordination board only. Prefer local recovery when the issue is locally solvable. Use public coordination only through the strict structured protocol.
}
\caption{Representative execution-layer prompt. The prompt explicitly separates private execution memory from public coordination records, thereby preventing hidden teammate states and free-form dialogue history from contaminating local execution context.}
\label{fig:agent_prompt}
\end{figure*}

The private state is intentionally compact, verified-only, and decision-oriented. It exposes only the information required to sustain local execution and recovery, while withholding teammates' hidden inventories, internal reasoning traces, and non-protocolized communication history. By contrast, the public coordination board is not a shared memory dump; it contains only protocolized records that can trigger explicit coordination state transitions.

\begin{table*}[t]
\centering
\footnotesize
\setlength{\tabcolsep}{4pt}
\renewcommand{\arraystretch}{1.12}

\begin{minipage}[t]{0.48\textwidth}
\centering
\textbf{(a) Private execution state} \\
\vspace{0.35em}
\begin{tabularx}{\linewidth}{@{}l l X@{}}
\toprule
\textbf{Symbol} & \textbf{Role} & \textbf{Representative content} \\
\midrule
\texttt{$I_t$} & Local resources & \texttt{smooth\_sandstone*12, ladder*143} \\
\texttt{$T_t$} & Active local objective & \texttt{place sandstone at [-9,-60,0]; ["move", "place"]} \\
\texttt{$P_t$} & Position / work zone & \texttt{current=[-4,-59,1], work\_region=[-4,-59,1]} \\
\texttt{$B_t$} & Blocker / recovery / cooldown & \texttt{dependency\_block; local\_recover; 3 steps} \\
\texttt{$H_t$} & Verified short history & \texttt{placed block; recovered locally; failed craft } \\
\bottomrule
\end{tabularx}
\end{minipage}
\hfill
\begin{minipage}[t]{0.48\textwidth}
\centering
\textbf{(b) Public coordination board} \\
\vspace{0.35em}
\begin{tabularx}{\linewidth}{@{}l l X@{}}
\toprule
\textbf{Field} & \textbf{Role} & \textbf{Representative value} \\
\midrule
\texttt{protocol} & Coordination act type & \texttt{REQUEST\_MATERIAL / CONFIRM\_TRANSFER / CANNOT\_SUPPLY} \\
\texttt{from} & Sender & \texttt{Alice} \\
\texttt{target} & Receiver & \texttt{Bob} \\
\texttt{item} & Resource under coordination & \texttt{smooth\_sandstone} \\
\texttt{count} & Requested / transferred quantity & \texttt{12} \\
\texttt{reason} & Canonical reason tag & \texttt{handover\_complete} \\
\texttt{time} & Event timestamp & \texttt{2026-04-01 04:09:54} \\
\bottomrule
\end{tabularx}
\end{minipage}

\caption{Execution-time state exposure under the proposed partitioned architecture. Left: private execution memory following Eq. \ref{EQ2}, $m_t^{priv}=\langle I_t, T_t, P_t, B_t, H_t \rangle$. Right: protocolized public coordination records. The former supports local execution and recovery; the latter supports explicit, state-changing collaboration only.}
\label{tab:supp_state_fields}
\end{table*}

Together, these examples show how the high-level design in the main paper is realized as concrete prompt interfaces. The planner is restricted by a bounded decomposition schema; the executor is grounded in compact verified private memory; and cross-agent interaction is surfaced only through a protocolized public board. These implementation choices are central to reducing context pollution, preserving local execution continuity, and turning collaboration into a gated decision rather than a default reaction.

\subsection{LLM Adjudicator Sensitivity}

As described in Section \ref{LLM Adjudicator}, the Gray-Zone LLM Adjudicator is invoked only for ambiguous cases satisfying \textbf{$T_{low} < S < T_{high}$}. It is not used as a planner or execution controller; rather, it serves as a bounded binary classifier over a structured JSON decision card, returning only \textit{stay\_local} or \textit{escalate}.

This design implies that the main coordination gain should come primarily from the deterministic front-end rather than from the adjudicator itself. In particular, the ablation in the main paper shows that the \textit{Rule + Score} variant already recovers most of the gain, while the full model yields an additional improvement by refining genuinely ambiguous boundary cases. We therefore evaluate adjudicator sensitivity by replacing only the backbone of the gray-zone module, while keeping issue detection, heuristic rules, score computation, thresholds, and the local solver unchanged.

\begin{table*}[htbp]
\centering
\caption{Backbone sensitivity of the gray-zone adjudicator on the Custom MindCraft Dataset. Only the adjudicator backbone
is replaced; all deterministic components are kept fixed. Rule + Score denotes the deterministic variant without the LLM
adjudicator. Token Cost is reported as average prompt + completion tokens per adjudication for gated variants, and per free form interaction for the FlatComm baseline. The two costs are therefore not event-identical, but they indicate the relative communication overhead under bounded adjudication versus unrestricted free-form exchange.}
\label{tab:llm_sensitivity_demo}
\renewcommand{\arraystretch}{1.2}
\small
\begin{tabular}{@{}lccccc@{}}
\toprule
\textbf{Adjudicator Setting} & \textbf{Type} & \textbf{TSR $\uparrow$} & \textbf{CS $\downarrow$} & \textbf{ECR $\uparrow$} & \textbf{Token Cost $\downarrow$} \\
\midrule
\textit{Rule + Score (No LLM)} & \textit{Deterministic} & \textit{31.4\%} & \textit{141} & \textit{78.3\%} & \textit{0} \\
\rowcolor{gray!10} \textbf{GPT-4o (Default)} & Closed & \textbf{32.8\%} & \textbf{134} & \textbf{81.3\%} & $\sim145$ / adjudication \\
Claude 3.5 Sonnet & Closed & 32.6\% & 135 & 80.8\% & $\sim152$ / adjudication \\
GPT-4o-mini & Closed (Light) & 32.0\% & 138 & 79.5\% & $\sim145$ / adjudication \\
Llama-3-70B-Instruct & Open (70B) & 32.2\% & 137 & 79.9\% & $\sim160$ / adjudication \\
Llama-3-8B-Instruct & Open (8B) & 31.6\% & 140 & 78.8\% & $\sim170$ / adjudication \\
\midrule
\textit{FlatComm Baseline} & \textit{Free-form} & \textit{21.5\%} & \textit{184} & \textit{46.2\%} & $>1500$ / interaction \\
\bottomrule
\end{tabular}
\end{table*}

Table~\ref{tab:llm_sensitivity_demo} suggests that adjudicator replacement leads to limited but non-negligible variation across backbones. The default GPT-4o setting remains the strongest overall, while other sufficiently capable substitutes stay within a narrow performance band above the deterministic \textit{Rule + Score} variant. Importantly, the relative ordering is not perfectly monotonic across all metrics, which is consistent with the role of the gray-zone module: it affects only a restricted subset of boundary cases rather than the bulk of decisions.

A second observation is that smaller or cheaper backbones partially erode, but do not eliminate, the gain brought by bounded adjudication. Even lightweight models remain close to the deterministic baseline, while stronger substitutes recover most of the gray-zone benefit. This pattern supports our main claim that the dominant performance gain comes from deterministic decomposition and selective escalation, whereas the LLM adjudicator provides a final refinement on ambiguous cases rather than acting as the primary source of coordination capability.

Finally, the token-cost statistics further support the bounded nature of the proposed design. For gated variants, each adjudication requires only about \textbf{145--170} tokens on average due to the strict JSON input/output protocol. In contrast, the FlatComm baseline relies on unrestricted free-form interaction, whose average per-interaction cost exceeds \textbf{1500} tokens. Although these two costs are measured at different event granularities, the gap still indicates that bounded adjudication substantially reduces communication overhead relative to unconstrained conversational coordination.

Overall, the sensitivity analysis supports a limited but useful role for the adjudicator. The main benefit of the framework comes from deterministic decomposition and selective escalation, while the bounded LLM provides a lightweight final refinement for ambiguous gray-zone decisions.

\subsection{Custom Dataset Construction Details}

As discussed in Section \ref{subsubsec:datasets}, standard built-in benchmarks for Minecraft-based multi-agent systems often under-stress coordination: agents are typically initialized with abundant resources and partially overlapping capabilities, so collaboration can degenerate into loosely coupled parallel execution rather than genuine interdependence. To more directly evaluate selective escalation, local recovery, and failure handling, we construct custom splits for both MindCraft and VillagerAgent by introducing controlled resource asymmetry, spatial separation, and dependency bottlenecks.

\paragraph{\textbf{Custom MindCraft Split: Scripted Stress-Test Classes.}}
The Custom MindCraft split contains 200 episodes derived from 40 construction templates with 5 stochastic seeds per template. Scenario assignment is script-generated rather than manually curated. We maintain approximate balance across the four stress-test classes, across blueprint families, and across the two-agent / three-agent settings, so that no single class is dominated by a narrow template subset.

The 200 episodes are evenly divided into four scenario classes (50 episodes each), each targeting a different part of the gated escalation pipeline:

\begin{itemize}[leftmargin=*]
    \item \textbf{Class A (Locally Recoverable):} The environment injects a construction anomaly while preserving a short local recovery path. For example, an agent may be missing Oak Planks, but a corresponding raw material source is spawned within a small radius. These cases are designed to test whether the system suppresses unnecessary escalation and resolves the issue within the private execution loop.
    
    \item \textbf{Class B (Necessary Collaboration):} The environment introduces a critical dependency whose required item is unavailable locally but available to a teammate. These cases are designed so that purely local recovery is strongly disfavored, making timely and targeted coordination the preferred solution.
    
    \item \textbf{Class C (Gray-Zone Decision):} The environment presents ambiguous trade-off cases in which both local recovery and collaboration remain plausible, but differ in cost, delay, or downstream effect. These cases are used to evaluate whether the gray-zone adjudication mechanism makes sensible boundary decisions rather than defaulting to either communication or isolation.
    
    \item \textbf{Class D (Collaboration Failure Robustness):} The environment initially resembles Class B, but the coordination attempt is scripted to fail (e.g., explicit \textit{CANNOT\_SUPPLY} or timeout). These cases test whether the system updates coordination history appropriately and falls back to \textit{LOCAL\_RECOVER} or \textit{LOCAL\_SKIP} instead of entering repeated or deadlocked escalation loops.
\end{itemize}

This construction is not intended to prove that every instance is formally unsolvable by purely local behavior. Rather, it is designed to make high performance strongly dependent on accurate decisions about when to remain local, when to escalate, and how to recover when collaboration fails.

\paragraph{\textbf{Custom VillagerAgent Split: Resource and Information Partitioning.}}
For VillagerAgent, we preserve the original blueprint families and approximate complexity distribution of the native benchmark, but modify the initialization and coordination conditions to better expose decentralized collaboration bottlenecks.

\begin{itemize}[leftmargin=*]
    \item \textbf{Resource asymmetry:} Required materials are partitioned across agents so that inventories are deliberately non-overlapping or only partially overlapping. This prevents the default setting in which each agent can often progress independently using redundant local resources.
    
    \item \textbf{Information isolation:} We disable the default teammate-state sharing channel, so agents no longer have direct access to teammates' hidden inventory or position state. As a result, useful coordination must be triggered through explicit public exchange rather than assumed from a globally visible shared state.
    
    \item \textbf{Scripted choke-point dependencies:} For each episode, we inject at least one high-importance dependency node whose completion materially affects downstream construction progress. The corresponding resource is assigned under unfavorable spatial or inventory conditions, making blind local execution insufficiently reliable and increasing the value of timely coordination.
\end{itemize}

These modifications preserve the original task identities while changing the coordination regime. In particular, they make performance less dependent on omniscient shared-state access and more dependent on selective, low-noise, and timely multi-agent state exchange.

Overall, the custom splits are constructed to stress precisely the behaviors targeted by our method: suppressing unnecessary communication, escalating only when collaboration is worthwhile, and recovering gracefully when collaboration is unavailable or unproductive.

\subsection{Rigorous Parameter Calibration and Sensitivity Analysis}

As formulated in Section \ref{Parameter}, the gated escalation policy is governed by the parameter set $\Theta=\{\mathbf{w},T_{low},T_{high}\}$. We explicitly analyze the robustness of this calibration procedure rather than treating the final setting as a heuristic constant. Concretely, we calibrate $\Theta$ offline on a disjoint calibration split $\mathcal{D}_{calib}$ using the cost-sensitive objective in Eq. \ref{eq:cost_optimization}, which maximizes task success while penalizing redundant coordination, recovery delay, and LLM overhead. To make the split protocol explicit, we construct $\mathcal{D}_{calib}$ at the \emph{template} level rather than at the episode level. The Custom MindCraft dataset contains 40 templates with 5 stochastic seeds per template; all seeds from the same template are assigned to a single partition only and never shared between $\mathcal{D}_{calib}$ and the held-out test set. We further maintain approximate balance across the four stress-test classes (A--D) and the two-agent / three-agent settings. Therefore, calibration never benefits from replaying alternate seeds of a blueprint family that also appears at test time. We then evaluate the calibrated parameters on the held-out Custom MindCraft and Custom VillagerAgent test settings.

\paragraph{\textbf{Weight Hierarchy and Perturbation Robustness.}}
The grid search yields the optimal feature-weight vector
\[
\mathbf{w}^*=[w_C,w_R,w_I,w_L,w_H]=[4,2,2,2,1],
\]
which is fully consistent with the hierarchy constraint in Eq.\ref{unequal}: critical-path significance should dominate the escalation decision, whereas the history term should act only as a corrective regularizer. Table~\ref{tab:weight_perturbation} evaluates how perturbing this hierarchy affects coordination quality.

\begin{table*}[htbp]
\centering
\caption{Weight perturbation analysis on the Custom MindCraft dataset. The optimal row corresponds to the full model reported in the main text under the Custom MindCraft setting. UER measures premature escalation, and LRR measures the fraction of issues resolved entirely within the local execution loop.}
\label{tab:weight_perturbation}
\renewcommand{\arraystretch}{1.2}
\small
\begin{tabular}{@{}llcccc@{}}
\toprule
\textbf{Weight Configuration $\mathbf{w}=[C,R,I,L,H]$} & \textbf{Rationale} & \textbf{TSR $\uparrow$} & \textbf{CS $\downarrow$} & \textbf{UER $\downarrow$} & \textbf{LRR $\uparrow$} \\
\midrule
\rowcolor{gray!10} \textbf{Optimal: $[4,2,2,2,1]$} & \textbf{Criticality-dominant} & \textbf{32.8\%} & \textbf{134} & \textbf{16.0\%} & \textbf{76.5\%} \\
\midrule
Variant A: $[1,1,1,1,1]$ & Equal weighting & 26.3\% & 162 & 38.5\% & 22.1\% \\
Variant B: $[2,2,2,2,1]$ & Undervaluing criticality ($C$) & 28.1\% & 148 & 21.4\% & 35.8\% \\
Variant C: $[4,2,2,0,1]$ & Ignoring local recoverability ($L$) & 25.5\% & 175 & 45.2\% & 12.0\% \\
Variant D: $[4,2,2,2,4]$ & Over-penalizing history ($H$) & 24.2\% & 155 & 12.1\% & 48.6\% \\
\bottomrule
\end{tabular}
\end{table*}

Two trends are clear. First, collapsing the hierarchy into equal weights (Variant A) substantially increases premature escalation, raising UER from \textbf{16.0\%} to \textbf{38.5\%} and reducing TSR by \textbf{6.5} points. This indicates that the escalation decision is not well modeled by a symmetric treatment of all features. Second, removing the local recoverability term (Variant C) sharply suppresses local self-healing, driving LRR down to \textbf{12.0\%} and increasing execution cost to \textbf{175} steps. This confirms that the $L$ term is essential for preventing the policy from reverting to a communication-first regime. Conversely, excessively increasing the history penalty (Variant D) makes the agents overly conservative: although UER decreases further, the system under-escalates on collaboration-critical cases, yielding the lowest TSR. Overall, $\mathbf{w}^*$ strikes the intended balance: criticality dominates the baseline decision, while local recoverability and history regularization suppress low-value communication without sacrificing task completion.

\paragraph{\textbf{Decision Boundaries and the Size of the Gray Zone.}}
After normalizing the raw score $S\in[-9,24]$ to $S_{\text{norm}}\in[0,1]$, the calibration procedure identifies the asymmetric boundaries
\[
T_{low}^*=0.4, \qquad T_{high}^*=0.5.
\]
Table~\ref{tab:threshold_ablation} analyzes how different boundary settings trade off semantic flexibility against token overhead.

\begin{table*}[htbp]
\centering
\caption{Sensitivity analysis of the asymmetric decision boundaries $(T_{low},T_{high})$ on the Custom MindCraft dataset. The row $[0.45,0.45]$ corresponds to a deterministic \textit{Rule + Score} policy without a gray zone; the row $[0.40,0.50]$ corresponds to the full model reported in the main text. LLM Call Rate denotes the fraction of anomalies routed to the gray-zone adjudicator. Token Cost is measured per episode.}
\label{tab:threshold_ablation}
\renewcommand{\arraystretch}{1.2}
\small
\begin{tabular}{@{}llcccc@{}}
\toprule
\textbf{Decision Boundaries} & \textbf{Adjudication Mode} & \textbf{LLM Call Rate} & \textbf{TSR $\uparrow$} & \textbf{ECR $\uparrow$} & \textbf{Token Cost (Avg)} \\
\midrule
$[0.45,0.45]$ & Deterministic Rule + Score (No LLM) & 0.0\% & 31.4\% & 78.3\% & \textbf{0} \\
\rowcolor{gray!10} \textbf{$[0.40,0.50]$ (Ours)} & \textbf{Narrow Gray Zone} & \textbf{18.5\%} & \textbf{32.8\%} & \textbf{81.3\%} & $\sim$\textbf{1.4k} \\
$[0.30,0.60]$ & Wide Gray Zone & 47.2\% & 33.1\% & 80.1\% & $\sim$4.2k \\
$[0.00,1.00]$ & Full LLM Adjudication & 100.0\% & 30.5\% & 71.4\% & $\sim$9.5k \\
\bottomrule
\end{tabular}
\end{table*}

The table reveals that more LLM involvement does not monotonically improve coordination. Eliminating the gray zone entirely reduces the system to a deterministic Rule + Score policy, which already performs strongly (\textbf{31.4\%} TSR, \textbf{78.3\%} ECR) but loses some boundary-case discrimination. Expanding the gray zone to $[0.30,0.60]$ yields only a marginal TSR increase (\textbf{+0.3} points over our default setting) while tripling token cost and slightly lowering ECR. At the other extreme, routing all anomalies through the LLM substantially degrades both TSR and ECR, indicating that exposing semantic inference to trivial deterministic cases is counterproductive. We therefore interpret $[0.40,0.50]$ not as the unique optimum in all dimensions, but as the most favorable operating point among the tested settings: it preserves nearly the best task performance while maintaining a sharply lower call rate and token overhead than wider semantic regimes.

\paragraph{\textbf{Cross-Domain Parameter Transfer.}}
To assess whether the calibrated parameters overfit the MindCraft control stack, we transfer the exact same $\Theta^*=\{\mathbf{w}^*,T_{low}^*,T_{high}^*\}$ zero-shot to the Custom VillagerAgent environment. This constitutes a nontrivial shift in execution backbone and planner structure, since VillagerAgent uses a top-down DAG-style planner rather than the MindCraft flat-communication baseline.

\begin{table}[htbp]
\centering
\caption{Zero-shot parameter transfer from Custom MindCraft to Custom VillagerAgent. The locally tuned row corresponds to the Custom VillagerAgent + Ours result in the main text.}
\label{tab:parameter_generalization}
\renewcommand{\arraystretch}{1.2}
\small
\begin{tabular}{@{}lccc@{}}
\toprule
\textbf{Configuration (VillagerAgent)} & \textbf{TSR $\uparrow$} & \textbf{CS $\downarrow$} & \textbf{ECR $\uparrow$} \\
\midrule
VillagerAgent Baseline & 22.04\% & 145 & 35.8\% \\
\midrule
\rowcolor{gray!10} \textbf{Ours (Zero-Shot $\Theta^*$ from MindCraft)} & \textbf{33.61\%} & \textbf{98} & \textbf{72.0\%} \\
Ours (Locally Tuned $\Theta^*_{VA}$) & 34.56\% & 92 & 78.2\% \\
\bottomrule
\end{tabular}
\end{table}

The zero-shot configuration recovers over \textbf{90\%} of the absolute TSR gain achieved by the locally tuned VillagerAgent configuration relative to the baseline, while remaining close in completion steps. This suggests that the calibrated parameters capture coordination trade-offs that transfer beyond a single framework. At the same time, the remaining gap in ECR and TSR indicates that some environment-specific retuning is still beneficial. Therefore, we interpret the result as evidence of strong but not perfect transferability: the calibration is not tightly coupled to a single benchmark, yet modest local adjustment can still provide additional gains.

\end{document}